# Interplay of ballistic and chemical effects in the formation of structural defects for Sn and Pb implanted silica

R.J. Green[a], A. Hunt[a], D.A. Zatsepin[b,c], D.W. Boukhvalov[d], J.A. McLeod[a], E.Z. Kurmaev[b], N.A. Skorikov[b], N.V. Gavrilov[e], A. Moewes[a]

[a]*Department of Physics and Engineering Physics, University of Saskatchewan, Saskatoon, Saskatchewan S7N 5E2, Canada*
[b]*Institute of Metal Physics, Russian Academy of Sciences-Ural Division, 18 Kovalevskoi Str., 620990 Yekaterinburg, Russia*
[c]*Ural Federal University, 19 Mira Str., 620002 Yekaterinburg, Russia*
[d]*School of Computational Sciences, Korea Institute for Advanced Study (KIAS), Hoegiro 87, Dongdaemun-Gu, Seoul, 130-722, Korean Republic*
[e]*Institute of Electrophysics, Russian Academy of Sciences-Ural Division, 620016 Yekaterinburg, Russia*

## Abstract

The electronic structures of Sn and Pb implanted $SiO_2$ are studied using soft X-ray absorption (XAS) and emission (XES) spectroscopy. We show, using reference compounds and *ab initio* calculations, that the presence of Pb-O and Sn-O interactions can be detected in the pre-edge region of the oxygen $K$-edge XAS. Via analysis of this interaction-sensitive pre-edge region, we find that Pb implantation results primarily in the clustering of Pb atoms. Conversely, with Sn implantation using identical conditions, strong Sn-O interactions are present, showing that Sn is coordinated with oxygen. The varying results between the two ion types are explained using both ballistic considerations and density functional theory calculations. We find that the substitution of Pb into Si sites in $SiO_2$ requires much more energy than substituting Sn in these same sites, primarily due to the larger size of the Pb ions. From these calculated formation energies it is evident that Pb requires far higher temperatures than Sn to be soluble in $SiO_2$. These results help explain the complex processes which take place upon implantation and determine the final products.

## 1. Introduction

In recent years, ion beam synthesis has become a widespread and valuable technique for the modification of the structural, optical, electronic, and magnetic properties of semiconductors and dielectrics. These properties are often modified through the formation of nano-sized embedded aggregates, which can lead to phenomena such as intense photoluminescence, altered band structures, nonlinear optical effects, and superparamagnetism [1, 2, 3]. The materials synthesized with this technique have attracted attention as candidates for promising new applications in fields such as optoelectronics.

Ion implantation is known to be a very versatile technique for material synthesis as one can embed almost any elemental ion within almost any host material in a very controlled manner. The success of the ion beam synthesis technique, however, relies heavily on the careful consideration of the many processes which take place upon ion beam irradiation of a material [4]. Variations in any experimental parameters such as ion type, implantation energy, target compound, temperature, etc., can lead to strong variations in the final products. An often successful approach to understanding or predicting the final products of implantation is termed the *two-step model* (see, for example, Refs. [2, 5]), which considers separately the effects of the high-energy physical step characterized by ballistic effects, and the subsequent low-energy chemical step based on thermodynamic considerations.

While the two-step model is often useful for predicting the physical and chemical state of an ion beam irradiated material, it is not always clear which step dominates for given implantation parameters. For example, high energy, high fluence implantation typically leads to more host damage. In such a case the ballistic effects are undoubtedly strong, but the severe rearrangement of the host lattice due to ballistic effects consequently provides a prime situation for chemical reordering. For low fluence, low energy implantation, disturbance to the lattice is small, so both ballistic and chemical effects are somewhat reduced, likely leading to only interstitial ions in the extreme case. Thus, at the current time, it is often difficult to tell *a priori* what form the final product will take and why.

In the current work we study the implantation of Pb and Sn ions into amorphous, bulk $SiO_2$, using periodically pulsed ion implantation and no thermal annealing. $SiO_2$ implanted with Sn [6, 7, 8] and Pb [9, 10] has shown promise in past studies regarding the formation of metallic or oxide nanoclusters and subsequent non-linear opti-

*Email address:* robert.green@usask.ca (R.J. Green)

cal properties and Coulomb blockade effects. A combination of X-ray absorption (XAS) and emission (XES) spectroscopy is used in the present work to closely study the structural and chemical effects of implantation. Through analysis of the spectra we identify a significant presence of Sn-O bonding, and a prominent lack of Pb-O bonding. The results are explained using the two step model as well as *ab initio* calculations of formation energies of defect structures. Finally, we shed light on aspects of the ion implantation process which will be important for future studies.

## 2. Experimental and Computational Details

The amorphous $SiO_2$ ($a$-$SiO_2$) samples used as targets for this study were 99.9% purity KB-type, plane-parallel glass plates measuring $1 \times 1$ cm$^2$ with surfaces of optical quality and thicknesses of ~0.7 mm. Separate samples were irradiated with Sn$^+$ and Pb$^+$ ions using a pulsed source with an ion beam current density of ~2 - 7 mA/cm$^2$, implantation energy of 30 keV, ion fluence of $5 \times 10^{16}$ cm$^{-2}$, and a pulse duration of 400 $\mu$s. The surface temperatures of the samples during Sn$^+$ and Pb$^+$ irradiation did not exceed 300 K due largely to the low currents used. No thermal annealing was performed after implantation. A high-purity $a$-$SiO_2$ sample was also left unirradiated for use as a reference material.

An additional reference material consisting of a PbO/$SiO_2$ mixture was synthesized via melting 99.9% purity PbO and $SiO_2$ in a corundum crucible under normal atmosphere. A ratio of 35:65 PbO:$SiO_2$ was used, and the mixture was kept at 1000-1300 °C for three hours. The samples were then baked at 350-400 °C for 20 minutes and slowly cooled. All samples under study and references were prepared in the Ural Federal University (Yekaterinburg, Russia).

The penetration depths for Sn$^+$ and Pb$^+$ ions, as well as damage profiles for the $SiO_2$ lattice, were estimated using the SRIM program [11]. The program uses a *Monte Carlo*-based binary collision approximation to model the interaction of the ions with the solid. Simulations were performed using the same materials and implantation energies implemented in the experiment, and the implantation of a total number of $10^5$ ions was simulated in each case to obtain adequately converged profiles.

The electronic structure calculations of pure PbO, SnO, and crystalline $SiO_2$ were performed using the *ab initio* WIEN2k code [12] which is based on the full-potential augmented plane-wave method with scalar-relativistic corrections. The modified Becke-Johnson variant of the generalized gradient approximation (GGA-MBJ) exchange-correlation functional was used [13], and the ground state density of states (DOS) was calculated in each case. The atomic sphere radii (the cutoff radius between plane waves and spherical harmonics) were chosen as $R_{MT} = 1.51$, 2.17, and 2.20 a.u. for the cations in $SiO_2$, PbO, and SnO, respectively, while for the O atoms they were 1.51, 2.17, and 1.75 a.u., respectively [14, 15]. They were chosen in such a way that the spheres were nearly touching. The Brillouin zone integrations were performed with a $k$-point grid of up to 800 points and $R_{MT}^{min} K_{max} = 7$ (the product of the smallest of the atomic sphere radii $R_{MT}$ and the plane wave cutoff parameter $K_{max}$) was used for the expansion of the basis set.

The formation energy modeling was carried out using the first-principles pseudopotential method as implemented in SIESTA [16]. All calculations were performed using the Perdew-Burke-Ernzerhof variant of the generalized gradient approximation (GGA-PBE) [17] for the exchange-correlation potential. The atomic positions were fully optimized until the inter-atomic force was less than 0.04 eV/Å. A $2 \times 2 \times 2$ supercell containing 108 (36 silicon and 72 oxygen) atoms was utilized. All calculations were carried out for an energy mesh cut off of 360 Ry and $k$-points of $4 \times 4 \times 4$ Monkhorst-Pack mesh [18]. To verify the performance of the pseudopotentials we performed an optimization of the crystal structure for bulk $\alpha$-quartz, silicon, tin and lead. In all cases the obtained deviation between experimental and calculated values of lattice parameters was less than 1.5%. The chemisorption energy has been calculated with the help of the standard formula $E_{chem} = (E_{Q+X} - (E_Q - E_{Si} + E_X))/n$, where $E_{Q+X}$ is the total energy of quartz supercell with an X impurity in a Si substitutional position, $E_{Si}$ and $E_X$ are the total energies per atom of silicon and impurity in a bulk phase, respectively, $n$ is a quantity of impurity atoms and $E_Q$ corresponds to the total energy of pure $\alpha$-quartz.

Soft X-ray emission spectra (XES) spectra were measured using Beamline 8.01 of the Advanced Light Source (ALS) at Lawrence Berkeley National Laboratory [19]. The beamline uses a Rowland circle type diffraction grating spectrometer with a 90° scattering angle and the incident radiation is plane-polarized in the horizontal direction. Soft X-ray absorption spectra (XAS) were measured using the Spherical Grating Monochromator (SGM) beamline of the Canadian Light Source (CLS) at the University of Saskatchewan [20]. Spectra were collected using the total fluorescence yield detection scheme, again with horizontal plane-polarized incident radiation.

## 3. Results

Figure 1 shows the simulated depth distributions of the Pb and Sn ions in $a$-$SiO_2$, using the parameters implemented in our experiment and displacement energies of 15 eV and 28 eV for the Si and O atoms of the host, respectively. These simulations provide a very good estimate for the ion distribution within a host for specific implantation parameters. We do note, however, that large fluences (such as those used in this work) may lead to non-negligible density changes in the target during implantation, resulting in an overall slightly shallower ion range than predicted. One can observe from the simulations that Pb and Sn are expected to penetrate similar distances into the $SiO_2$ on average. The ion range calculated for Pb implantation is 20.1 nm, whereas that for Sn is 20.8 nm. Additionally, we see that for the case of Pb the distribution is more spatially localized than that for Sn. We note that the scales for the upper and lower panels have the same values to help show that the Sn concentration does not peak as high as for Pb, and that the ions are distributed in a more gradual nature. The spread of the distributions can be quantified via the calculated average straggling (square root of the variance). For Pb implantation the straggling is 3.4 nm, whereas for Sn it is 5.2 nm.

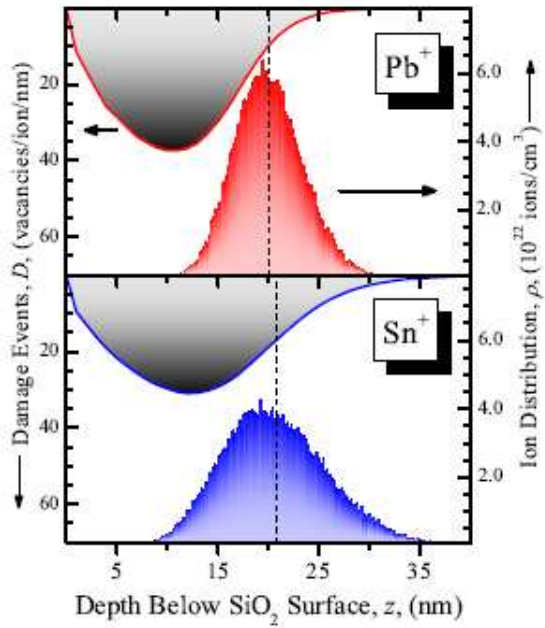

Figure 1: (Color online) Ion depth distributions and host lattice damage of Pb$^+$ and Sn$^+$ implanted in SiO$_2$. The profiles were simulated using the SRIM program [11] with the experimental implantation parameters given in Section 2.

The simulated amount of damage events (Si and O vacancies generated in the $a$-SiO$_2$ either by the implanted ions or recoiling target atoms) are also shown in Figure 1 for each case. Note these plots are shown with reversed y-axes for clarity. It must be stressed that again changes in the material during the process are not taken into account for each successive implanted ion in the SRIM simulation, so these damage profiles are not expected to correspond exactly to experimental results. Additionally, the simulations are performed for a temperature 0 K, whereas at room temperature implantation it is known that many damage events are removed through self-annealing. Lastly, we recall that a pulsed ion beam approach was used in this work, which is expected to reduce the amount of damage to the host compared to steady state implantation [21].

With these caveats considered, the simulations nonetheless provide a good estimate of how the damage to the host should change for the two different ion cases. Similar to the ion distributions, we see that the damage events for the Pb case are more localized and intense near the surface region, while for Sn the damage is spread more evenly and deeper into the $a$-SiO$_2$.

The overlap of the final ion distribution with the implantation region can also be visualized in Figure 1. There is a noticeably larger overlap of the regions for the case of Sn implantation, due to the spread out nature of the Sn regions as compared to Pb. Since this is an important observation that will be considered in Section 4, we can quantify in each case the overlap of damage region with the final ion distribution using the overlap integral

$$I = \frac{\int_0^\infty \sqrt{\rho(z) D(z)} dz}{\sqrt{\int_0^\infty \rho(z) dz}\sqrt{\int_0^\infty D(z) dz}} \quad (1)$$

where $I$ is the overlap integral (essentially a Bhattacharyya coefficient), $\rho$ is the density of the implanted ions at depth $z$ below the surface, and $D$ is the number of damage events at depth $z$. The unitless integral is normalized between zero (no overlap) and one (complete overlap), so that it may be directly compared between cases. Computing the overlap integrals for each case, we obtain $I_{Sn} = 0.776$ and $I_{Pb} = 0.538$, confirming that the overlap of the ion distribution with the damage region is indeed significantly larger for the Sn implantation case. This is perhaps expected: Pb ions should inflict greater damage than Sn ions in a more localized region near the surface due to their larger radius and mass, *and* the Pb ions should be able to travel further into the SiO$_2$ than Sn *after* losing energy to damage events due to their larger mass.

Figure 2 displays the oxygen $K$-edge XAS and XES for the reference $a$-SiO$_2$ and the Pb-containing samples in the upper panel, while the spectra for the Sn-containing samples are shown (again with the $a$-SiO$_2$ for direct comparison) in the bottom panel. While the XES reveals a somewhat similar valence band structure for the various materials, there are strong differences in the conduction band nature as probed by XAS. In particular, the large electronic band gap of $a$-SiO$_2$ (~9 eV) leads to an XAS spectrum which is quite high in energy. The spectra of SnO, SnO$_2$, PbO, and PbO:SiO$_2$ conversely, contain prominent features at lower energies (~531 - 535 eV). The spectra of the implanted samples fall somewhere between the extremes: Sn implanted $a$-SiO$_2$ shows a prominent peak in the pre-edge of a spectrum (at ~532.5 eV) which is otherwise similar to pure $a$-SiO$_2$, while Pb implanted $a$-SiO$_2$ is almost identical to that of pure $a$-SiO$_2$, with only minor spectral weight in the pre-edge region. Note that the similarity of the XAS for Pb implanted $a$-SiO$_2$ and the reference $a$-SiO$_2$ has already been briefly noted in an earlier study which revealed a lack of PbO$_4$ tetrahedra in the samples [22].

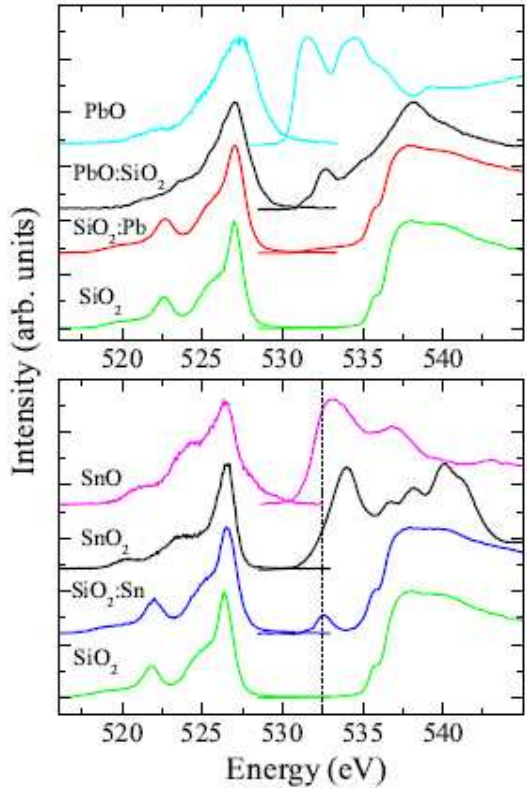
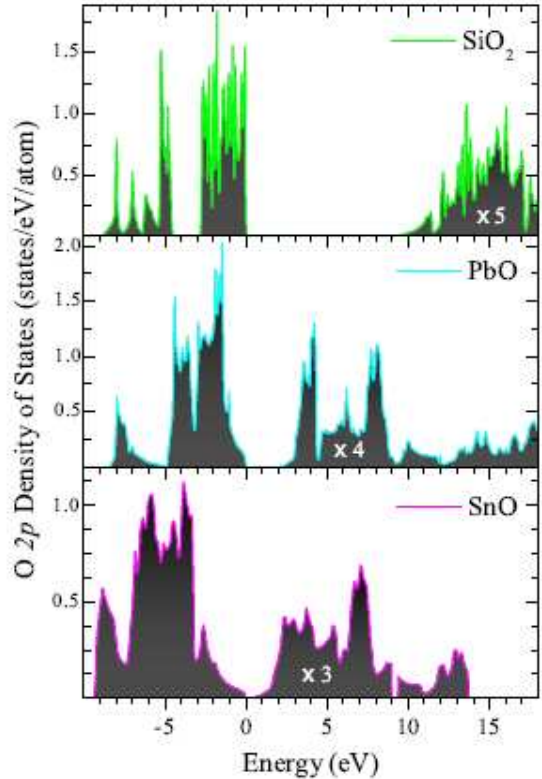

Figure 2: (Color online) Oxygen $K$-edge X-ray absorption (right) and emission (left) spectra for implanted and reference samples. Samples containing Pb are shown in the upper panel, while Sn-containing samples are shown in the bottom panel.

Figure 3: (Color online) Calculated O $2p$ partial densities of states for $SiO_2$, PbO, and SnO. The conduction band states have been scaled by the indicated values for display purposes.

To verify the nature of the observed features in the spectra of Figure 2, O $2p$ partial densities of states for $SiO_2$, PbO, and SnO were calculated and are shown in Figure 3. This is essentially what is probed with the XES and XAS, though the XAS spectra measure a DOS which is perturbed by the $1s$ core hole created during the measurement. The $SiO_2$ calculation was approximated with $\alpha$-quartz due to computational limitations, but should be an adequate approximation for this application as $a$-$SiO_2$ and $\alpha$-quartz exhibit very similar XAS and XES spectra, the main difference being $\alpha$-quartz has sharper, more distinct features [23] due to the increased order. The basic structure of the occupied states is in agreement with the measured XES - we see a strong higher energy band paired with a lower intensity, lower energy band as per the XES. The unoccupied states also support the results of the measured XAS. For $SiO_2$, the conduction band is quite high in energy compared to the cases of PbO and SnO. Thus, a very useful test for the presence of Pb-O and Sn-O bonding within a $a$-$SiO_2$ matrix is the identification of the lower energy conduction band states.

Finally in Figure 4, we show the Si $L_{2,3}$ XES spectra of the implanted samples, along with the reference $a$-$SiO_2$. Since the spectra of $a$-$SiO_2$ and Si are significantly different [24], using this technique one can detect the formation of Si-Si bonds, and thus the relative amount of damage to the $SiO_2$ lattice due to the implantation [25]. Here we see the spectra are all very similar, but the Pb implanted $SiO_2$ shows some slight variations. Notably, there is increased weight between the two main peaks (denoted by the arrow in the figure), which is an indication of the Si-Si bonds as mentioned above. For $SiO_2$:Sn there is only a very minor increase in spectral weight in this region. These results can confirm the more intense, localized damage near the surface for Pb implantation, as this would yield more noticeable effects in the XES than the spread out nature of the Sn damage. Note that at these energies, the X-rays should just be able to probe the entire implantation region, but there will be a preference toward the surface due to the attenuation of the incoming and outgoing X-rays. Thus, these results support the prediction from the simulations in Figure 1 that the Pb implantation leads to more localized damage closer to the surface than Sn. In fact, the XES results suggest that the actual differences in damage distributions might be slightly more pronounced than the SRIM calculations, as we see very little change in the XES for Sn implantation compared to that of Pb.

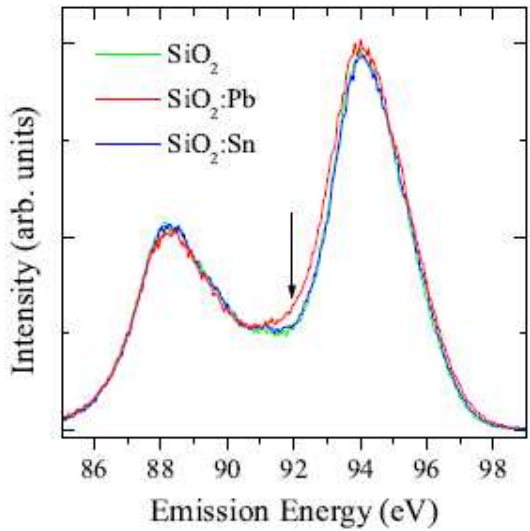

Figure 4: (Color online) Silicon $L_{2,3}$-edge X-Ray Emission spectra for implanted and reference $SiO_2$.

## 4. Discussion

From the XAS spectra in Figure 2, we see that there is minimal interaction between the implanted Pb ions and the O atoms in the $a$-$SiO_2$. For the $PbO$:$SiO_2$ solution on the other hand, there is pronounced spectral weight in the 531-535 eV region, indicating strong Pb-O interactions (as expected). While one likely would not see such strong spectral weight from Pb-O bonding in the $SiO_2$:Pb due to the lower relative concentration of Pb, it provides an indication of the type of feature one would expect to see. Thus, it is evident that in this case the Pb ions interact minimally with the O atoms of the lattice (either through the formation of oxide clusters or as single substitutional or interstitial atoms), and instead they likely aggregate into metallic clusters, as found in previous studies [9, 10]. The gradual onset in the pre-edge of the $SiO_2$:Pb XAS spectrum could be due to interactions between the surfaces of the Pb clusters and the O atoms of the lattice, or simply due to damage in the $SiO_2$. For $SiO_2$:Sn, on the other hand, we see significantly stronger spectral weight in the pre-edge XAS which indicates there are significantly more Sn-O interactions than Pb-O interactions.

Previous studies of Pb implanted $SiO_2$ have found (before thermal annealing) Pb clusters [9, 10] as well as Pb-O interactions [10]. Note that both of these studies used a much higher implantation energy (~300 keV) than the current study (30 keV) which leads to ion depths of ~100 nm rather than the ~20 nm here. In these studies, Pb clusters formed preferentially under lower fluences (entirely Pb clusters at $5 \times 10^{15}$ cm$^{-2}$ for [9] and increasing Pb-O contributions as fluence was increased from 1 to $6 \times 10^{16}$ cm$^{-2}$ in [10]). The explanation for the increasing Pb-O interactions in [10] was suggested to be increased damage as the fluence was increased. This helps to explain the lack of Pb-O interactions in the present study, as the lower implantation energy leads to lower damage in the $SiO_2$, and thus, fewer dangling bonds with which the Pb ions may interact.

Earlier thermodynamics considerations using the two-step model predicted both the formation of Pb aggregates as well as Pb-O interactions [2]. The former seems to happen in almost all cases, but the latter often does not, as in the present study. This might be due to variations in the first (ballistic) step of the process. We have seen that variations in implantation energy (which lead to different damage conditions) lead to different results regarding the formation of Pb-O interactions. The low energy case studied here, combined with a pulsed ion beam, leads to an overall gentler process which appears to eliminate the formation of Pb-O interactions, whereas the higher energy case does not.

Previous studies of Sn implanted $SiO_2$ have often found the formation of $SnO_2$ nanoparticles [26, 27], but typically only after thermal oxidation is performed. Accordingly, samples with Sn nanoparticles have been formed via vacuum annealing after ion implantation [28]. In the present study, no annealing has been performed, so the possibility of oxide aggregates is somewhat reduced. One might still expect Sn-O interactions, however, via interstitial Sn ions throughout the implantation region or Sn ions at Si sites within the less damaged regions of the $SiO_2$ lattice. As explained above with reference to the XAS spectra, we do indeed see prominent Sn-O interactions. Returning to this spectrum (Figure 2), it is evident that it is not simply a sum of SnO or $SnO_2$ with $SiO_2$, suggesting that, as expected without annealing, the Sn-O interactions are not simply from the formation of SnO or $SnO_2$ nanoclusters. Instead, it seems likely the Sn has been incorporated into the $SiO_2$ lattice.

To determine why the cases of Sn and Pb implantation lead to different results, we can consider more closely both steps of the two step model. First, considering the ballistic step, we have seen from the implantation simulations and the Si XES that Pb implantation leads to more localized damage near the surface as well as a more localized ion distribution, whereas Sn implantation leads to damage and an ion distribution which are more spread into the bulk of the $SiO_2$ and accordingly overlap somewhat. The overlap integrals described in Section 3 showed that for the Sn case the ions tend to end up in locations where there are vacancies and broken bonds present, whereas for Pb more ions end up at positions of less damage. This helps to explain why Sn is more apt to bond with O atoms than Pb is for the current implantation conditions.

| Impurity | (S)ingle/(P)air | Supercell Expansion (%) | $E_{chem}$ (eV) |
|---|---|---|---|
| Pb | S | 3.0 | 9.56 |
| Pb | P | 2.0 | 8.84 |
| Sn | S | 1.2 | 4.91 |
| Sn | P | 2.0 | 4.95 |

Table 1: Calculated formation energies and structural expansion properties of $SiO_2$ with selected defects.

Considering now the thermodynamic processes as the second step, we have employed a DFT pseudopotential calculations of formation energies for substitutional defects (Pb and Sn substituting at Si sites in $SiO_2$). Existing two step model predictions [2] which have been referred to above, consider rather the formation of Sn and Pb oxide clusters for the ion-oxygen interactions. Since our results differ slightly from these predictions, it seems necessary to consider the more advanced calculations of Sn and Pb incorporation into the $SiO_2$ structure. Note that while this is an approximation to the physical case where some damage is present, it should provide a useful method to compare the two cases. We have optimized atomic positions and calculated total energies for $SiO_2$ with single and pair substitutional impurities of both Sn and Pb at Si sites (with the pairs placed at nearest neighbouring Si positions). The configurations studied correspond to 4 and 8% Pb and Sn concentrations for single and pair substitutions, respectively. The results of the calculations are summarized in Table 1. Notably, the larger Pb ions necessitate a larger expansion of the lattice upon substitution, and correspondingly have much higher formation energies than Sn ions. This agrees with our results, as it suggests that Sn is much more likely to diffuse into $SiO_2$ lattice positions than Pb, as we have observed in our experiment. Note also the lower formation energy for the Pb case when two impurities are incorporated in neighbouring Si sites (8.84 eV) as opposed to a single impurity (9.56 eV). This provides insight into why the Pb aggregates into metallic formations, as the lower energies for neighbouring Pb might catalyze the formation of the clusters.

Thus, the explanation for the different behaviors of Pb and Sn upon implantation in $SiO_2$ under identical conditions can be summarized as:

- Pb ions with the current implantation parameters lead to more localized damage away from the implantation region. Thus the ions are situated in an area of reduced dangling bonds. Additionally, Pb is only soluble in $SiO_2$ at very high temperatures, so ions are not likely to diffuse into single positions (substitutional or interstitial) in the $SiO_2$ lattice. This leads primarily to the metallic aggregation of the Pb ions.

- Sn ions under the current implantation parameters are more spread out and situated largely in the damage region. Thus there are opportunities (vacancies, dangling bonds) for the Sn to be incorporated into the lattice. Additionally, Sn has better solubility in $SiO_2$ than Pb. This leads to significant Sn-O interactions rather than Sn aggregation.

With these results obtained for the present study, some key opportunities for future studies arise. Through systematic variation of the implantation parameters (fluence, energy, target temperature), one could test the parameter boundaries which lead to different final products. For example, it is likely that for fixed fluence and temperature, the implantation energy could be varied to alter the overlap of the damage region and final ion locations.

## 5. Conclusion

To conclude, an electronic structure study of $Sn^+$ and $Pb^+$ ion-implanted, amorphous $SiO_2$ has been performed. The presence of Sn-O interactions was revealed, whereas Pb aggregation was found rather than Pb-O interactions. These results were explained with careful consideration of both the ballistic and chemical steps of the implantation process, in the vein of the existing two step model. Ion ranges and damage profiles for the ballistic step and density functional theory calculations of formation energies for the chemical step both support our experimental observations. The driving force behind the different results for each ion is found primarily to be the ion size (ionic radius for chemical considerations, and overall mass and size for ballistic considerations). These results shed light on materials that can be synthesized from Sn and Pb implantation, as well as on implantation processes in general.

## 6. Acknowledgments

The present study was carried out under the support of the Natural Sciences and Engineering Research Council of Canada (NSERC), the Canada Research Chair program, and the Ural Division of the Russian Academy of Sciences (Project No. 12-I-2-2040). Assistance from the staffs of the Advanced Light Source and Canadian Light Source is gratefully acknowledged. The Advanced Light Source is supported by the Director, Office of Science, Office of Basic Energy Sciences, of the U. S. Department of Energy under Contract No. DE-AC02-05CH11231. The Canadian Light Source is supported by NSERC, the National Research Council (NSC) Canada, the Canadian Institute of Health Research (CIHR), the Province of Saskatchewan, Western Economic Diversification Canada, and the University of Saskatchewan.

# 7. References

[1] A. Meldrum, R. Haglund, L. Boatner, C. White, Nanocomposite materials formed by ion implantation, Adv. Mater. 13 (2001) 1431.
[2] E. Cattaruzza, Quantum-dot composite silicate glasses obtained by ion implantation, Nucl. Instrum. Methods Phys. Res., Sect. B 169 (2000) 141–155.
[3] A. L. Stepanov, Nonlinear optical properties of implanted metal nanoparticles in various transparent matrixes:a review, Rev. Adv. Mater. Sci. 27 (2011) 115–145.
[4] B. Ziberi, F. Frost, M. Tartz, H. Neumann, B. Rauschenbach, Importance of ion beam parameters on self-organized pattern formation on semiconductor surfaces by ion beam erosion, Thin Solid Films 459 (2004) 106–110.
[5] R. Bertoncello, A. Glisenti, G. Granozzi, G. Battaglin, F. Caccavale, E. Cattaruzza, P. Mazzoldi, Chemical interactions in titanium-implanted and tungsten-implanted fused-silica, J. Non-Cryst. Solids 162 (1993) 205–216.
[6] Y. Takeda, T. Hioki, T. Motohiro, S. Noda, Large 3rd-order optical nonlinearity of tin microcrystallite-doped silica glass formed by ion-implantation, Appl. Phys. Lett. 63 (1993) 3420–3422.
[7] Y. Takeda, T. Hioki, T. Motohiro, S. Noda, T. Kurauchi, Nonlinear-optical properties of $Sn^+$ ion-implanted silica glass, Nucl. Instrum. Methods Phys. Res., Sect. B 91 (1994) 515–519.
[8] A. Nakajima, T. Futatsugi, H. Nakao, T. Usuki, N. Horiguchi, N. Yokoyama, Microstructure and electrical properties of Sn nanocrystals in thin, thermally grown $SiO_2$ layers formed via low energy ion implantation, J. Appl. Phys. 84 (1998) 1316–1320.
[9] F. Luce, F. Kremer, S. Reboh, Z. Fabrim, D. Sanchez, F. Zawislak, P. Fichtner, Aging effects on the nucleation of Pb nanoparticles in silica, J. Appl. Phys. 109 (2011) 014320.
[10] R. Magruder, D. Henderson, S. Morgan, R. Zuhr, Optical spectra of Pb implanted fused silica, J. Non-Cryst. Solids 152 (1993) 258–266.
[11] J. Ziegler, J. Biersack, M. Ziegler, The Stopping and Range of Ions in Matter: Electronic Manual for SRIM Program, 2009.
[12] P. Blaha, K. Schwarz, G. Madsen, D. Kvasnicka, J. Luitz, WIEN2k, An Augmented Plane Wave + Local Orbitals Program for Calculating Crystal Properties: Electronic Book, Techn. Universitat Wien, Austria, 2001.
[13] F. Tran, P. Blaha, Accurate band gaps of semiconductors and insulators with a semilocal exchange-correlation potential, Phys. Rev. Lett. 102 (2009) 226401.
[14] P. Boher, P. Garnier, J. Gavarri, A. Hewat, Quadratic PbO-alpha monoxide 1. description of the ferroelastic structural transition, J. Solid State Chem. 57 (1985) 343–350.
[15] B. Will, M. Bellotto, W. Parrish, M. Hart, Crystal-structures of quartz and magnesium germanate by profile analysis of synchrotron-radiation high-resolution powder data, J. Appl. Crystallogr. 21 (1988) 182–191.
[16] J. Soler, E. Artacho, J. Gale, A. Garcia, J. Junquera, P. Ordejon, D. Sanchez-Portal, The SIESTA method for ab initio order-N materials simulation, J. Phys.: Condens. Matter 14 (2002) 2745–2779.
[17] J. Perdew, K. Burke, M. Ernzerhof, Generalized gradient approximation made simple, Phys. Rev. Lett. 77 (1996) 3865–3868.
[18] H. Monkhorst, J. Pack, Special points for brillouin-zone integrations, Phys. Rev. B 13 (1976) 5188–5192.
[19] J. Jia, T. Callcott, J. Yurkas, A. Ellis, F. Himpsel, M. Samant, J. Stöhr, D. Ederer, J. Carlisle, E. Hudson, L. Terminello, D. Shuh, R. Perera, First experimental results from IBM/TENN/TULANE/LLNL/LBL undulator beamline at the Advanced Light Source, Rev. Sci. Instrum. 66 (1995) 1394–1397.
[20] T. Regier, J. Krochak, T. Sham, Y. Hu, J. Thompson, R. Blyth, Performance and capabilities of the Canadian Dragon: The SGM beamline at the Canadian Light Source, Nucl. Instrum. Methods Phys. Res., Sect. A 582 (2007) 93–95.
[21] D. Zatsepin, E. Kurmaev, I. Shein, V. Cherkashenko, S. Shamin, S. Cholakh, Effect of high doses on the Si $L_2,L_3$ X-ray emission spectra of silicon implanted with iron ions under steady-state conditions, Phys. Solid State 49 (2007) 75–81.
[22] D. Zatsepin, A. Hunt, A. Moewes, E. Kurmaev, N. Gavrilov, I. Zhidkov, S. Cholakh, $Pb^+$ implanted $SiO_2$ probed by soft x-ray emission and absorption spectroscopy, J. Non-Cryst. Solids 357 (2011) 3381–3384.
[23] B. Gilbert, B. Frazer, F. Naab, J. Fournelle, J. Valley, G. De Stasio, X-ray absorption spectroscopy of silicates for in situ, sub-micrometer mineral identification, Am. Mineral. 88 (2003) 763–769.
[24] T. Sham, S. Naftel, P.-S. Kim, R. Sammynaiken, Y. Tang, I. Coulthard, A. Moewes, J. Freeland, Y.-F. Hu, S. Lee, Electronic structure and optical properties of silicon nanowires: A study using x-ray excited optical luminescence and x-ray emission spectroscopy, Phys. Rev. B 70 (2004) 045313.
[25] D. Zatsepin, R. Green, A. Hunt, E. Kurmaev, N. Gavrilov, A. Moewes, Structural ordering in a silica glass matrix under Mn ion implantation, J. Phys.: Condens. Matter 24 (2012) 185402.
[26] P. Kuiri, H. Lenka, J. Ghatak, G. Sahu, B. Joseph, D. Mahapatraa, Formation and growth of $SnO_2$ nanoparticles in silica glass by Sn implantation and annealing, J. Appl. Phys. 102 (2007) 024315.
[27] M. Tagliente, V. Bello, G. Pellegrini, G. Mattei, P. Mazzoldi, M. Massaro, $SnO_2$ nanoparticles embedded in silica by ion implantation followed by thermal oxidation, J. Appl. Phys. 106 (2009) 104304.
[28] R. Giulian, F. Kremer, L. Araujo, D. Sprouster, P. Kluth, P. Fichtner, A. Byrne, M. Ridgway, Shape transformation of Sn nanocrystals induced by swift heavy-ion irradiation and the necessity of a molten ion track, Phys. Rev. B 82 (2010) 113410.